%% file: main.tex
\documentclass[conference]{IEEEtran}
\IEEEoverridecommandlockouts
\usepackage{cite}
\usepackage{amsmath,amssymb,amsfonts}
\usepackage{algorithmic}
\usepackage{graphicx}
\usepackage{comment}
\graphicspath{ {./Images/} }
\usepackage{textcomp}
\usepackage{xcolor}
\usepackage{float}
\usepackage{subcaption}

\DeclareMathOperator*{\argmin}{arg\,min}

\def\BibTeX{{\rm B\kern-.05em{\sc i\kern-.025em b}\kern-.08em
    T\kern-.1667em\lower.7ex\hbox{E}\kern-.125emX}}

\begin{document}

\title{WIP: Federated Learning for Routing in Swarm Based Distributed Multi-Hop Networks
}

\author{\IEEEauthorblockN{Martha Cash,
Joseph Murphy, Alexander Wyglinski}
\IEEEauthorblockA{Department of Electrical and Computer Engineering,
Worcester Polytechnic Institute\\
Worcester, MA\\
\{mcash, jrmurphy, alexw\}@wpi.edu}
}

\maketitle

\begin{abstract}
Unmanned Aerial Vehicles (UAVs) are a rapidly emerging technology offering fast and cost-effective solutions for many areas, including public safety, surveillance, and wireless networks. However, due to the highly dynamic network topology of UAVs, traditional mesh networking protocols, such as the Better Approach to Mobile Ad-hoc Networking (B.A.T.M.A.N.), are unsuitable. To this end, we investigate modifying the B.A.T.M.A.N. routing protocol with a machine learning (ML) model and propose implementing this solution using federated learning (FL). This work aims to aid the routing protocol to learn to predict future network topologies and preemptively make routing decisions to minimize network congestion. We also present an FL testbed built on a network emulator for future testing of the proposed ML aided B.A.T.M.A.N. routing protocol. 
\end{abstract}


\section{Introduction}\label{sec:Intro}
\input{Sections/Introduction.tex}

\section{System Model}\label{sec:SysMod}
\input{Sections/SystemModel.tex}

\section{Proposed Solution}\label{sec:ProposedSln}
\input{Sections/ProposedSolution.tex}

\section{Preliminary Results}\label{sec:PreLim}
\input{Sections/PreliminaryResults.tex}

\section{Conclusion}\label{sec:FutureWrk}
\input{Sections/FutureWork.tex}

\section*{Acknowledgment}
This research was sponsored by the DEVCOM Analysis Center and was accomplished  under  Cooperative  Agreement  Number  W911NF-22-2-0001. The views and conclusions contained in this document are  those  of  the  authors  and  should  not  be  interpreted  as representing the official policies, either expressed or implied, of  the  Army  Research  Office  or  the  U.S.  Government.  The U.S.  Government  is  authorized  to  reproduce  and  distribute reprints  for  Government  purposes,  notwithstanding  any copyright notation herein.

\bibliography{IEEEabrv,references}

\end{document}

%% file: Sections/Introduction.tex
Unmanned Aerial Vehicles (UAVs) are a rapidly developing technology that has been used in numerous applications, including transportation, traffic control, surveillance, search and rescue, and disaster management \cite{b3}. Although UAV technology has many advantages, numerous challenges still need to be addressed to implement networking protocols for UAV-based infrastructures \cite{b2}. For one, UAV networks are highly dynamic. They do not have a consistent topology making communication, control, and path planning protocols designed for less dynamic mobile ad-hoc networks (MANETs) less effective \cite{b3}. These challenges motivate a need for routing protocols catered to UAVs. Ideally, these protocols should be simple, have low overhead, and not require extensive global topology knowledge  \cite{b4}. Further, because of the dynamic nature of UAVs, these protocols should make decisions based on the expected network topology rather than just the current state of the network. These requirements make AI-based routing protocols for UAVs appealing.

AI-based routing protocols are not a new area of research. For example, in \cite{b5} a supervised feed-forward neural network (FFNN) was proposed to learn network traffic history to adaptively route packets and improve heterogeneous network control. Using the Open Shortest Path First (OSPF) routing algorithm, the model input was an array representing the number of packets that were forwarded through each node in the network, while the model's output was the interface to forward the packet along. Simulation results demonstrated the effectiveness of the proposed FFNN approach and outperformed the OSPF baseline. Similarly, in \cite{b6}, Boltzmann machines (RBM) were proposed where the input was characterized as the traffic pattern observed at each router. Like \cite{b5}, this approach outperforms the baseline OSPF routing algorithm. Finally, the research presented in \cite{b7} proposed a neural network (NN) trained at each link in the network, and the output of the NN was the likelihood of successful packet delivery if the packet was forwarded along that link. The authors propose using buffer capacity, number of successful packet transfers, and node popularity.

All of these approaches focus on applying common machine learning (ML) techniques to the AI-based routing protocol problem in homogeneous networks, which will not work UAV networks. However, an emerging ML technique, federated learning (FL) \cite{b21}, has yet to be explored as a solution for heterogeneous networks. In this work, we propose an FL-based approach to the AI-based routing protocol problem, specifically for UAV swarms. We narrow our focus to the Better Approach to Mobile Ad-hoc Networking (B.A.T.M.A.N.) protocol \cite{b8}, propose modifying the algorithm using a NN model, and characterize the dataset necessary for this problem. Finally, we present an FL emulation environment built on the Extendable Mobile Ad-hoc Network Emulator (EMANE) \cite{EMANE} that will be used for testing the proposed solution. 

The remainder of this paper is organized as follows: Section \ref{sec:SysMod} provides an overview of the system model. Section \ref{sec:ProposedSln} presents the proposed solution and describes FL in more detail. Section \ref{sec:PreLim} shows the preliminary results of FL setup in a network emulator and NN model, and Section \ref{sec:FutureWrk} discusses conclusions and future research directions.

%% file: Sections/SystemModel.tex
\begin{figure*}[t]
    \centering
    \includegraphics[width=\textwidth]{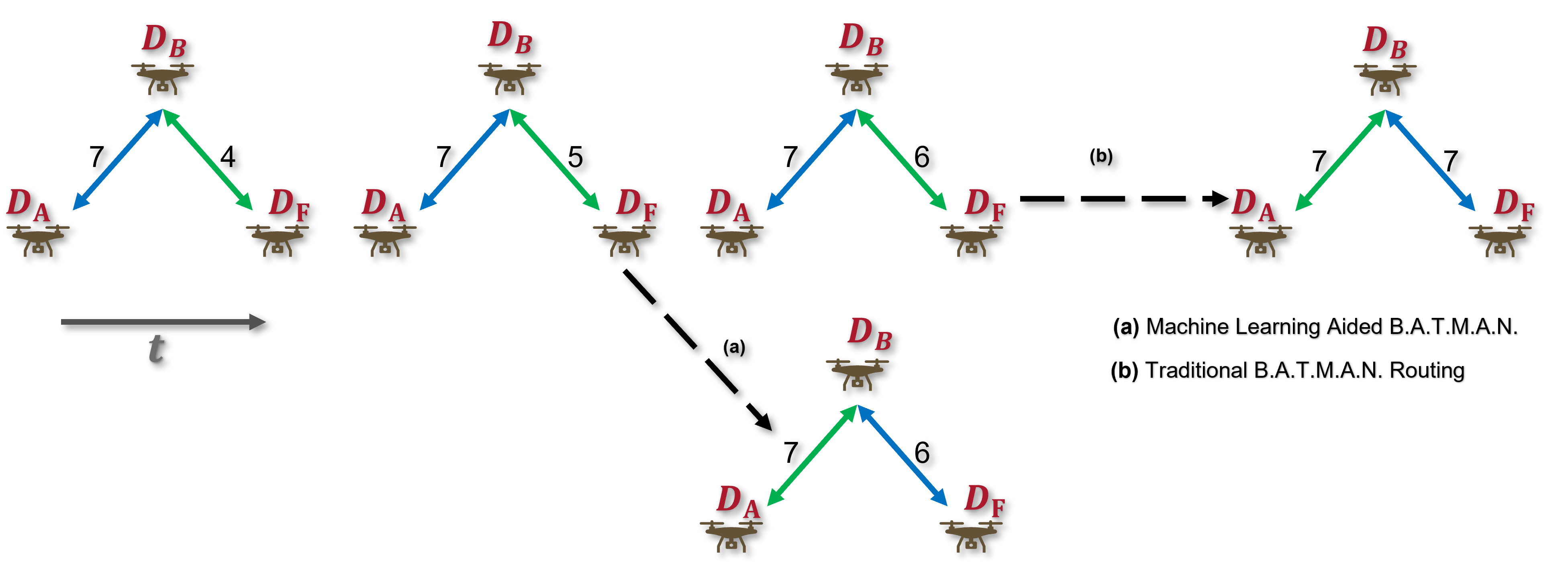}
    \caption{A two node UAV network. At each time step, the link cost between $D_{B}$ and $D_{F}$ increases while the link cost between $D_{B}$ and $D_{F}$ is constant. The B.A.T.M.A.N. routing protocol will continue to select the route with the lower link cost, branch $(b)$. The proposed solution, branch $(a)$, would preemptively switch routes to manage network congestion and select an alternate route despite being unfavorable at the current time step.}
    \label{fig:SysMod}
\end{figure*} 

The initial system, shown in Fig.~\ref{fig:SysMod}, is a simple two-node UAV network, which will be extended to a multi-node heterogeneous UAV network in future work. The network employs the B.A.T.M.A.N. routing protocol for communication between UAVs. The aim of this study, illustrated branch $(a)$ in Fig.~\ref{fig:SysMod}, is to demonstrate the feasibility of an ML-aided B.A.T.M.A.N. protocol to improve network congestion by predicting when to switch routes, even if switching to a route with a higher cost is not immediately beneficial. 

\subsection{B.A.T.M.A.N. Protocol}
As a baseline, the network uses the B.A.T.M.A.N routing protocol~\cite{batman}. B.A.T.M.A.N. was designed to address the challenges of routing in mobile ad-hoc networks (MANETs), such as frequent topology changes and the lack of a central authority to coordinate routing. Rather than maintain information about the global network topology, B.A.T.M.A.N. only requires nodes to maintain information about the best next hop to its immediate neighbors. The network is flooded with originator messages (OGMs). OGMs routed through good paths are received by nodes quicker than those transmitted on poor quality routes, informing the nodes in the network which immediate neighbor has the best route to transmit across. The routing tables are configured by selecting the best next hop to the originator node \cite{b11}.

However, there are a few drawbacks to the B.A.T.M.A.N. routing algorithm. For one, if the network contains a substantial number of nodes, B.A.T.M.A.N. can generate a large amount of overhead, as each node must re-broadcast the OGM to its neighbors. This can lead to increased network congestion and reduced overall efficiency. Additionally, in scenarios where the source or destination of the packet is in motion, B.A.T.M.A.N. can suffer from higher delays, which is undesirable if the network topology is highly dynamic. Finally, the drawback we focus on in this work is that B.A.T.M.A.N. is a threshold-based routing protocol, demonstrated in Fig.~\ref{fig:SysMod}. As a result, the node will always choose the next hop with the best route, even if conditions on the current best route are degrading. Waiting to change routes until the threshold is met can cause bottlenecks in the network \cite{b12}.

%% file: Sections/ProposedSolution.tex
We propose two solutions. First, we integrate federated learning (FL) to a wireless drone swarm network and propose an ML model and data set for enhancing the B.A.T.M.A.N. routing protocol. Next, we demonstrate an FL simulation environment built on the EMANE emulation environment, which will be integrated with the proposed ML model in future work to investigate a larger UAV network, and to introduce movement among the UAVs, since this will be a feature compensated for by the ML model. 

\subsection{Machine Learning Model \& Dataset}\label{sec:MLBat}
We consider a supervised learning approach for this work. The objective of supervised learning is to learn a mapping, or function approximation, $\hat{\mathbb{F}}(\mathbf{x},\mathbf{y})$, between a set of samples, $x_{i} \in X$, and their labels, $y_{i} \in Y$, where $X$ and $Y$ are the sample space and label space, respectively. Ideally, $\hat{\mathbb{F}}(x,y)$ takes a set of new samples, $\mathbf{x^{*}}$ and produces the correct label, $\mathbf{y^{*}}$. The quality of the mapping is determined by the loss function, $L(y^{*}_{i}, \hat{y}_{i})$, where $y^{*}_{i}$ is the true label of the new sample, and $\hat{y}_{i}$ is the output of $\hat{\mathbb{F}}({x^{*}_{i}}, \cdot)$ \cite{b18}. An accurate function approximation is quantified by a low loss value. 
 
Since the B.A.T.M.A.N. routing protocol does not maintain a history of route conditions (\textit{i.e.} link cost, throughput), we need to modify the B.A.T.M.A.N. algorithm to include a memory element. The model should learn a history of the prior link costs for each route, and the route the node selected. These requirements make the long short-term memory (LSTM) model, a type of recurrent neural network (RNN) that is designed to learn long-term dependencies in sequential data, appropriate for this task \cite{b14}.

The input to the LSTM model is a two dimensional array of the history of the link cost at each neighboring route from $D_{B}$, (see Fig.~\ref{fig:SysMod}). However, this approach can be extended to $n$ dimensions for $n$ many neighbors in a more complex network. The corresponding labels are a history of the selected route for transmission.  Instead of feeding the entire history of the network to the model, we implement a windowing technique. For example, if the window size is set to 4, then four prior time steps are fed into the model for training. We can treat this as a classification problem and use the binary cross entropy loss function since we are training the LSTM to select which of the two routes from $D_{B}$ to transmit across. 

\subsection{Federated Learning Approach}
Traditional machine learning approaches are centralized, meaning a model is trained on a central dataset that is collected and stored in a central location. However, this approach assumes all devices on the network have the same computational capabilities and network resources, which is often not the case for UAV networks \cite{b2}. As a result, a FL-based distributed ML technique is employed in this research where training of a global model is performed on data distributed across many UAVs in various locations. For generality, we assume there are $J$ UAVs in the network. Each UAV for $j \in J$ observes a unique dataset $\mathbf{x_{j}} = \{x_{j1}, x_{j2}, \ldots, x_{jN}\}$. Since we use a supervised learning approach, we assume a single input sample, $X_{jn}$ corresponds to a single output $y_{jn} \in \{y_{j1}, y_{j2}, \ldots, y_{jN}\}$. The sets $\mathbf{x}_{j}$ and $\mathbf{y}_{j}$ are used to train the local ML model at each UAV. Let $\mathbf{w}_{j} \in \mathbf{w}$ denote the corresponding model parameters at the $j^{th}$ UAV. Then, the FL objective function can be employed, which is defined as~\cite{b19}:

\begin{equation}\label{eq:FLObj}
\argmin_{\mathbf{w} \in \mathbb{R}^{d}} F(\mathbf{w}) = \frac{1}{N} \sum_{j}^{J}\sum_{n=1}^{N_{j}}f(\mathbf{w}_{j}, x_{jn}, y_{jn}). 
\end{equation}

We solve (\ref{eq:FLObj}) via the following steps: First, all UAVs are initialized with random parameters. Each UAV trains on its respective training sets, $x_{jn}$ and $y_{jn}$. After one epoch of training, the FL parameters at the $j^{th}$ UAV are sent to a central server. Once all parameters are received, the central server aggregates the parameters according to the following expression:

\begin{equation}\label{eq:fedAvg}
w_{global} = \frac{1}{J}\sum_{j=1}^{J} \mathbf{w_{j}}
\end{equation}

The global parameters, $w_{global}$, are sent back to the UAVs, and the training process repeats for $K$ epochs, or until $F(\mathbf{w})$ has converged to the optimal parameters, $w^{*}$. Fig.~\ref{fig:FLEMANE} summarizes this approach. 

\subsection{Federated Learning in EMANE}\label{sec:FLinEMANE}
\begin{figure}[!t]
    \centering
    \includegraphics[width = \linewidth]{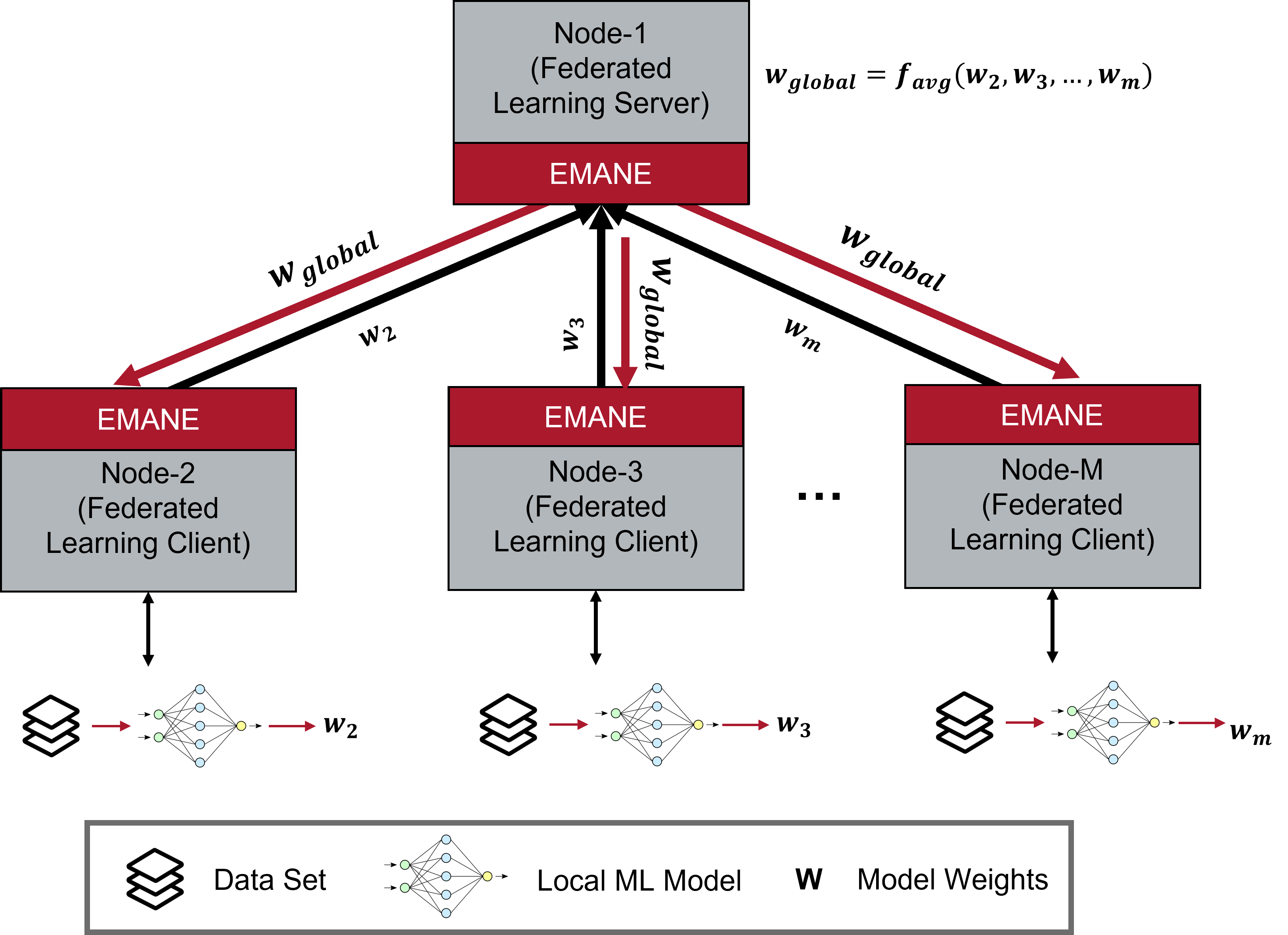}
    \caption{Federated Learning Setup in EMANE}
    \label{fig:FLEMANE}
\end{figure}

This works also aims to build an FL emulation environment, with the future goal of integrating the emulator with the proposed ML model to test the feasibility of the proposed solution. A system diagram of the emulator is shown in Fig.~\ref{fig:FLEMANE}. At the core of the emulator is EMANE, which allows for the creation of Network Emulation Modules (NEMs) to model different radio interface types. In turn, these can be incorporated into a real-time emulation running in a distributed environment and allow the direct integration of standard software, such as PyTorch, for handling ML tasks. 

For the results in this work, we construct a simulator with three NEMs, similar to the model setup in Fig.~\ref{fig:SysMod}. One NEM is designated as the central server. The remaining NEMs carry out the FL task. However, this could be generalized to $M$ nodes, see Fig.~\ref{fig:FLEMANE}. 

%% file: Sections/PreliminaryResults.tex
\subsection{CIFAR10 Baseline}
\begin{figure}[!t]
    \centering
    \includegraphics[width = \linewidth]{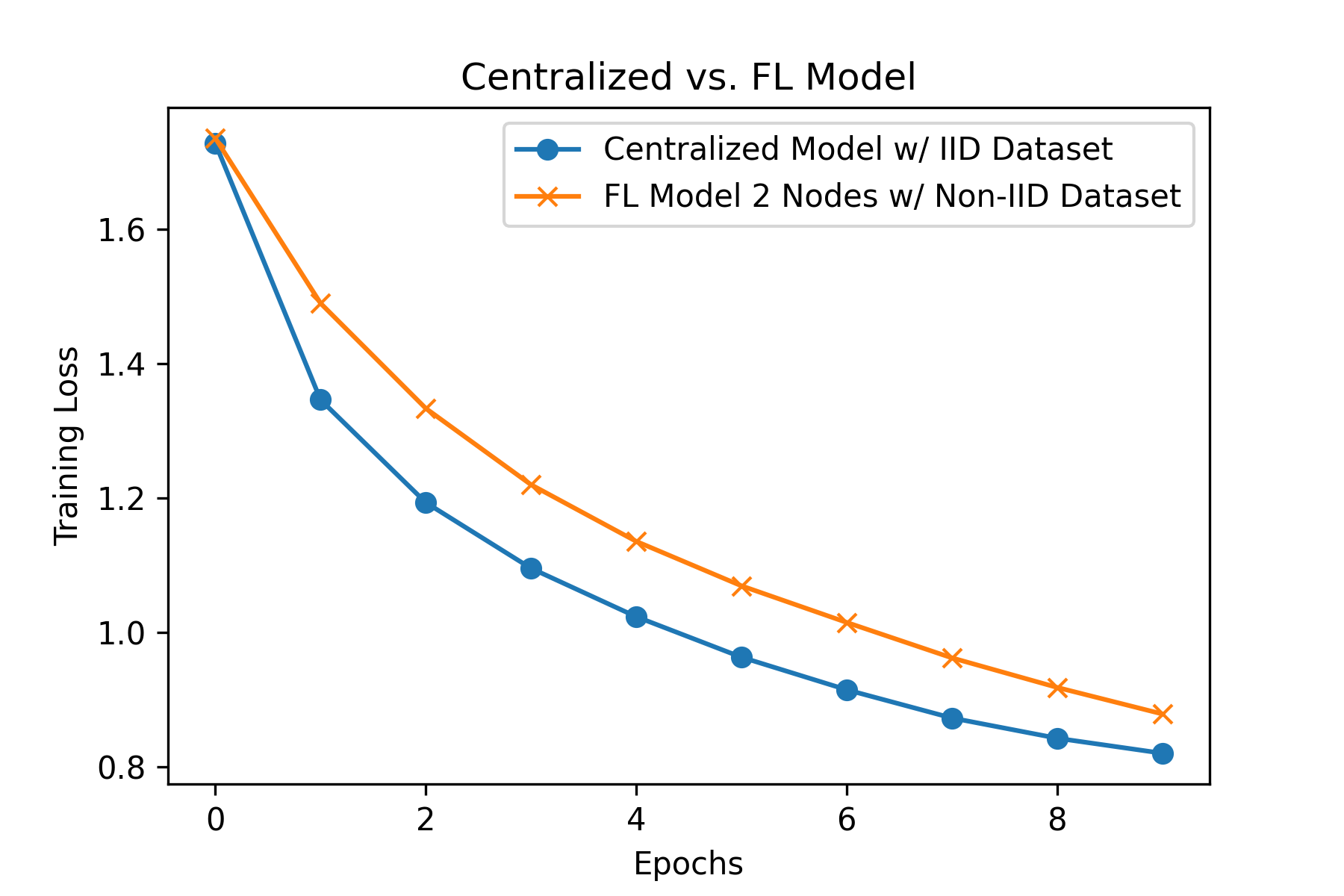}
    \caption{Federated Learning in EMANE Results: FL Model Approximates Centralized Model Performance}
    \label{fig:FLEMANERes}
\end{figure}

We use the CIFAR10 dataset \cite{b20} to test the performance of the network emulator proposed in Section \ref{sec:FLinEMANE} and compare the results to a centralized ML approach. To accurately model a distributed UAV environment, we randomly sample the CIFAR10 dataset so that each NEM in the emulator is not training on an identical dataset. We train the two edge nodes for 10 epochs. The results are presented in Fig.~\ref{fig:FLEMANERes}, demonstrating that the FL emulator ('x' curve) approximated the traditional ML approach ('o' curve). Because the nodes are not training on an identical dataset, the FL training loss curve should approximate the behavior of the centralized ML model training curve, but should not identically match the curve. The results presented in Fig.~\ref{fig:FLEMANERes} indicate the system setup proposed in Section \ref{sec:FLinEMANE} works and is suitable for future integration with the ML model presented in Section \ref{sec:MLBat} in addition to expanding the number of NEMs included in the emulation. 

\subsection{Machine Learning Aided B.A.T.M.A.N.}
\begin{figure}[!t]
\centering
    \includegraphics[width = \linewidth]{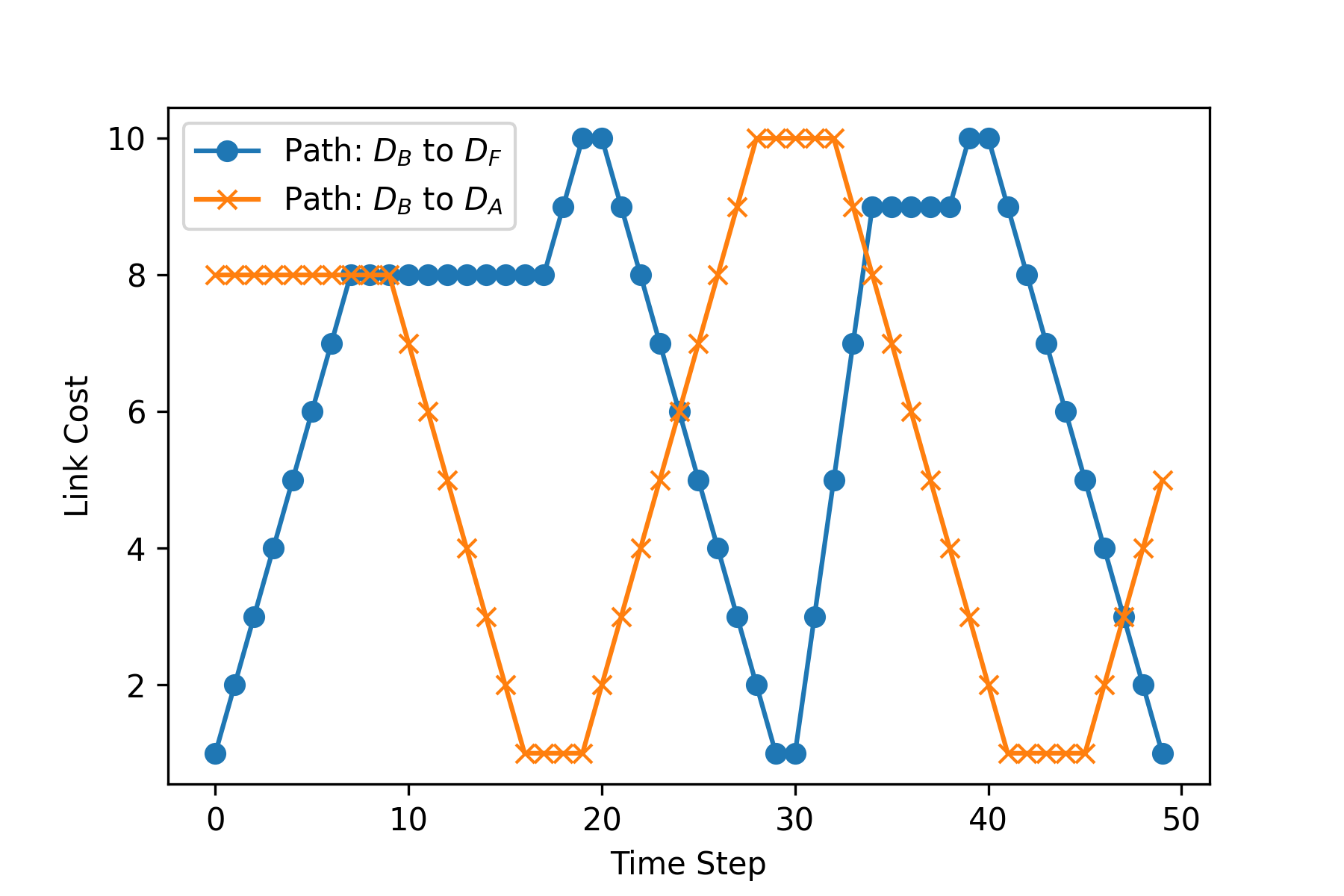}
    \caption{Demonstration of Hand Generated Training Set}
    \label{fig:TrainSet}
\end{figure}

We attempt to hand generate a simple testing and training set consisting of 50 time samples of link costs across the two routes branching from $D_{B}$ as demonstrated in Fig.~\ref{fig:TrainSet}. We construct the LSTM to have two recurrent layers. The input sequence length is set to four. We use a batch size of five and train the model for ten epochs using the binary cross entropy loss function and the ADAM optimizer with a learning rate of 0.01. Initial testing accuracy results show the LSTM has an 100\% classification accuracy. This is due to the memoryless nature of the hand generated data set. Therefore, in this current state, it is not possible to accurately estimate the practicality of the proposed model.

%% file: Sections/FutureWork.tex
This paper proposed a machine learning (ML) solution implemented via federated learning (FL) to modify the B.A.T.M.A.N. routing protocol for unmanned aerial vehicle networks. We presented a FL testbed built on the network emulator EMANE and used the CIFAR10 dataset to compare the FL testbed to a traditional centralized ML approach. The baseline results proved the testbed works as a proof of concept for future work. We also propose modifying the B.A.T.M.A.N. algorithm using a long short term memory model. However, current results do not accurately reflect the viability of this modification. Therefore, future work will need to generate a dataset from a simulation of the UAV swarm using EMANE to assess this approach better. 

%% file: main.bbl
\begin{thebibliography}{10}
\providecommand{\url}[1]{#1}
\csname url@samestyle\endcsname
\providecommand{\newblock}{\relax}
\providecommand{\bibinfo}[2]{#2}
\providecommand{\BIBentrySTDinterwordspacing}{\spaceskip=0pt\relax}
\providecommand{\BIBentryALTinterwordstretchfactor}{4}
\providecommand{\BIBentryALTinterwordspacing}{\spaceskip=\fontdimen2\font plus
\BIBentryALTinterwordstretchfactor\fontdimen3\font minus
  \fontdimen4\font\relax}
\providecommand{\BIBforeignlanguage}[2]{{%
\expandafter\ifx\csname l@#1\endcsname\relax
\typeout{** WARNING: IEEEtran.bst: No hyphenation pattern has been}%
\typeout{** loaded for the language `#1'. Using the pattern for}%
\typeout{** the default language instead.}%
\else
\language=\csname l@#1\endcsname
\fi
#2}}
\providecommand{\BIBdecl}{\relax}
\BIBdecl

\bibitem{b3}
A.~Rovira-Sugranes, A.~Razi, F.~Afghah, and J.~Chakareski, ``A review of
  ai-enabled routing protocols for uav networks: Trends, challenges, and future
  outlook,'' \emph{Ad Hoc Networks}, vol. 130, p. 102790, 2022.

\bibitem{b2}
\BIBentryALTinterwordspacing
L.~Gupta, R.~Jain, and G.~Vaszkun, ``Survey of important issues in uav
  communication networks,'' \emph{Commun. Surveys Tuts.}, vol.~18, no.~2, p.
  1123–1152, apr 2016. [Online]. Available:
  \url{https://doi.org/10.1109/COMST.2015.2495297}
\BIBentrySTDinterwordspacing

\bibitem{b4}
M.~Khaledi, A.~Rovira-Sugranes, F.~Afghah, and A.~Razi, ``On greedy routing in
  dynamic uav networks,'' in \emph{2018 IEEE International Conference on
  Sensing, Communication and Networking (SECON Workshops)}.\hskip 1em plus
  0.5em minus 0.4em\relax IEEE, 2018, pp. 1--5.

\bibitem{b5}
N.~Kato, Z.~M. Fadlullah, B.~Mao, F.~Tang, O.~Akashi, T.~Inoue, and
  K.~Mizutani, ``The deep learning vision for heterogeneous network traffic
  control: Proposal, challenges, and future perspective,'' \emph{IEEE Wireless
  Communications}, vol.~24, no.~3, pp. 146--153, 2017.

\bibitem{b6}
B.~Mao, Z.~M. Fadlullah, F.~Tang, N.~Kato, O.~Akashi, T.~Inoue, and
  K.~Mizutani, ``Routing or computing? the paradigm shift towards intelligent
  computer network packet transmission based on deep learning,'' \emph{IEEE
  Transactions on Computers}, vol.~66, no.~11, pp. 1946--1960, 2017.

\bibitem{b7}
D.~K. Sharma, S.~K. Dhurandher, I.~Woungang, R.~K. Srivastava, A.~Mohananey,
  and J.~J. P.~C. Rodrigues, ``A machine learning-based protocol for efficient
  routing in opportunistic networks,'' \emph{IEEE Systems Journal}, vol.~12,
  no.~3, pp. 2207--2213, 2018.

\bibitem{b21}
B.~McMahan, E.~Moore, D.~Ramage, S.~Hampson, and B.~A. y~Arcas,
  ``Communication-efficient learning of deep networks from decentralized
  data,'' in \emph{Artificial intelligence and statistics}.\hskip 1em plus
  0.5em minus 0.4em\relax PMLR, 2017, pp. 1273--1282.

\bibitem{b8}
A.~Klein, L.~Braun, and F.~Oehlmann, ``Performance study of the better approach
  to mobile adhoc networking (b.a.t.m.a.n.) protocol in the context of
  asymmetric links,'' in \emph{2012 IEEE International Symposium on a World of
  Wireless, Mobile and Multimedia Networks (WoWMoM)}, 2012, pp. 1--7.

\bibitem{EMANE}
\BIBentryALTinterwordspacing
U.~N.~R. Laboratory, ``Extendable mobile ad-hoc network emulator (emane).''
  [Online]. Available:
  \url{\url{https://www.nrl.navy.mil/Our-Work/Areas-of-Research/Information-Technology/NCS/EMANE/}}
\BIBentrySTDinterwordspacing

\bibitem{batman}
\BIBentryALTinterwordspacing
M.~Lindner, S.~Eckelmann, S.~Wunderlich, M.~Hundeb\o{}ll, A.~Quartulli, and
  L.~L\"{u}ssing, ``The b.a.t.m.a.n. project.'' [Online]. Available:
  \url{http://www.open-mesh.org/}
\BIBentrySTDinterwordspacing

\bibitem{b11}
\BIBentryALTinterwordspacing
B.~Sliwa, S.~Falten, and C.~Wietfeld, ``Performance evaluation and optimization
  of {B.A.T.M.A.N.} {V} routing for aerial and ground-based mobile ad-hoc
  networks,'' \emph{CoRR}, vol. abs/1901.02298, 2019. [Online]. Available:
  \url{http://arxiv.org/abs/1901.02298}
\BIBentrySTDinterwordspacing

\bibitem{b12}
D.~Seither, A.~König, and M.~Hollick, ``Routing performance of wireless mesh
  networks: A practical evaluation of batman advanced,'' in \emph{2011 IEEE
  36th Conference on Local Computer Networks}, 2011, pp. 897--904.

\bibitem{b18}
A.~Valadarsky, M.~Schapira, D.~Shahaf, and A.~Tamar, ``A machine learning
  approach to routing,'' \emph{arXiv preprint arXiv:1708.03074}, 2017.

\bibitem{b14}
S.~Hochreiter and J.~Schmidhuber, ``Long short-term memory,'' \emph{Neural
  computation}, vol.~9, no.~8, pp. 1735--1780, 1997.

\bibitem{b19}
T.~Zeng, O.~Semiari, M.~Mozaffari, M.~Chen, W.~Saad, and M.~Bennis, ``Federated
  learning in the sky: Joint power allocation and scheduling with uav swarms,''
  in \emph{ICC 2020 - 2020 IEEE International Conference on Communications
  (ICC)}, 2020, pp. 1--6.

\bibitem{b20}
A.~Krizhevsky, G.~Hinton \emph{et~al.}, ``Learning multiple layers of features
  from tiny images,'' 2009.

\end{thebibliography}
